\documentclass{ecai} 

\usepackage{latexsym}
\usepackage{amssymb}
\usepackage{amsmath}
\usepackage{amsthm}
\usepackage{booktabs}
\usepackage{enumitem}
\usepackage{graphicx}
\usepackage{color}
\usepackage{bm, cases}
\usepackage[dvipsnames]{xcolor}
\usepackage{ragged2e}

\newtheorem{theorem}{Theorem}

\newtheorem{proposition}[theorem]{Proposition}

\newtheorem{definition}{Definition}

\newtheorem{kq}{Key Question}
\newtheorem{obs}{Observation}

\newcommand{\BibTeX}{B\kern-.05em{\sc i\kern-.025em b}\kern-.08em\TeX}

\definecolor{nr}{RGB}{187, 213, 151}
\definecolor{opra}{RGB}{246, 216, 143}
\definecolor{oprb}{RGB}{127, 184, 222}
\definecolor{bprss}{RGB}{208, 137, 151}
\definecolor{bprop}{RGB}{190, 151, 197}

\begin{document}


\begin{frontmatter}


\paperid{4528} 


\title{Generative AI as Digital Representatives in Collective Decision-Making: A Game-Theoretical Approach}


\author[A,B]{\fnms{Kexin}~\snm{Chen} 
	\footnote{Email: kexinchen2@link.cuhk.edu.cn}
}
\author[A,B]{\fnms{Jianwei}~\snm{Huang} 
	\thanks{Corresponding author. Email: jianweihuang@cuhk.edu.cn}
}
\author[A]{\fnms{Yuan}~\snm{Luo} 
	\footnote{Email: luoyuan@cuhk.edu.cn}
}

\address[A]{School of Science and Engineering, The Chinese University of Hong Kong, Shenzhen}
\address[B]{Shenzhen Institute of Artificial Intelligence and Robotics for Society, Shenzhen Key Laboratory of Crowd Intelligence Empowered Low-Carbon Energy Network, and CSIJRI Joint Research Centre on Smart Energy Storage}


\begin{abstract}
	Generative Artificial Intelligence (GenAI) enables digital representatives to make decisions on behalf of team members in collaborative tasks, but faces challenges in accurately representing preferences. 
	While supplying GenAI with detailed personal information improves representation fidelity, feasibility constraints make complete information access impractical. 
	We bridge this gap by developing a game-theoretic framework that models strategic information revelation to GenAI in collective decision-making. 
	The technical challenges lie in characterizing members' equilibrium behaviors under interdependent strategies and quantifying the imperfect preference learning outcomes by digital representatives. 
	Our contribution includes closed-form equilibrium characterizations that reveal how members strategically balance team decision preference against communication costs. 
	Our analysis yields an interesting finding: Conflicting preferences between team members drive competitive information revelation, with members revealing more information than those with aligned preferences. 
	While digital representatives produce aggregate preference losses no smaller than direct participation, individual members may paradoxically achieve decisions more closely aligned with their preferences when using digital representatives, particularly when manual participation costs are high or when GenAI systems are sufficiently advanced.
	
\end{abstract}

\end{frontmatter}


\section{Introduction}	\label{sec: intro}
The rapid development of Generative Artificial Intelligence (GenAI) is accelerating its transition from a passive tool to an active participant in decision-making processes \cite{collins2024building}. 
Recent breakthroughs exemplify this evolution: Manus \cite{manus} independently completes complex online tasks, Google AI Co-Scientist \cite{googleco} leverages Gemini 2.0 to evaluate and refine scientific hypotheses, and as \citet{zou2025rise} highlighted, GenAI-powered agents serve as valuable teammates for human clinicians. 
These advances demonstrate that GenAI's advanced reasoning capabilities show great promise as digital representatives in collective decision-making \cite{jarrett2023language, tessler2024ai}.

However, a fundamental technical challenge persists: Merely prompting GenAI to role-play based on demographic information yields biased and inaccurate representations \cite{lee2024language, orlikowski2025beyond, petrov2024limited, wang2025limits}. 
Recent studies suggested that revealing richer individual data, such as social media histories or personality assessments, can enhance the accuracy of digital representatives \cite{anthis2025llm, park2024generative}. 
This approach has shown promise across various domains (e.g., \cite{bakker2022fine, burton2024large, estornell2024multi, horton2023large, jarrett2023language, tessler2024ai}).
For example, \citet{jarrett2023language} studied consensus-building by fine-tuning pre-trained GenAI to represent diverse human preferences. 
\citet{bakker2022fine} helped people find a common place in collective decision-making by training reward models to learn people's diverse opinions and then using GenAI to generate candidate agreements.
These studies advance the practical realization of digital representatives.

Concurrently, theoretical explorations have begun to investigate GenAI's role as a digital representative in settings such as auctions, mechanism design, and assignment problems (e.g., \cite{duetting2024mechanism, huang2025accelerated, soumalias2025llm}). 
For instance, \citet{huang2025accelerated} integrated GenAI into auction frameworks, where GenAI-powered representatives inferred bidders' preferences and interacted with the auction mechanism on their behalf.
\citet{duetting2024mechanism} examined a token auction model where advertisers encoded preferences into GenAI outputs for aggregated content generation. 
However, these studies primarily focus on high-level system design, assuming that GenAI can fully capture and represent members' preferences. 
This assumption rarely holds in practice, as fully articulating preferences over all alternatives would require infinite cost and effort, leading to imperfect digital representation.
This critical gap between theoretical models and practical constraints motivates our first key question:
\begin{kq}
	How to quantify the preference learned by the digital representatives?
\end{kq}

\begin{figure*}[t]
	\centering
	\includegraphics[width=.9\textwidth]{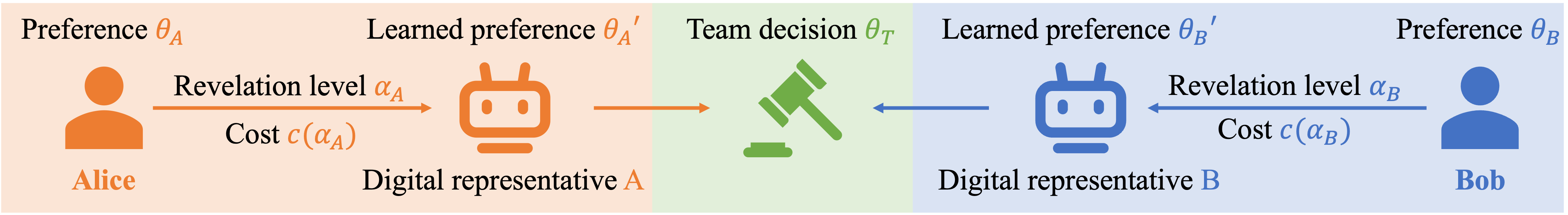}
	\caption{Illustration of the digital representative game.}
	\label{fig: 2-model}
\end{figure*}

Our formulation extends Castro et al.'s \cite{castro2024human} Bayesian framework for single-user information revelation to a multi-agent collective decision-making setting (Figure \ref{fig: 2-model}). 
This extension introduces three significant technical challenges: 
\textit{1)} Strategic interdependence among members' information revelation decisions, 
\textit{2)} Complex trade-offs between team decision preference and communication costs, 
and \textit{3)} Equilibrium analysis under heterogeneous preferences relative to GenAI's prior. 
These challenges require novel game-theoretic analysis techniques, leading to our second research question:

\begin{kq}
	How do members strategically interact with GenAI as digital representatives in collective decision-making?
\end{kq}

We develop closed-form equilibrium characterizations and analyze how system parameters influence these equilibria. 
Our key finding reveals a counter-intuitive competitive information revelation dynamic: Members with conflicting preferences reveal more information than those with aligned preferences, as they compete to prevent others from shifting the team decision against their preferences.

After characterizing equilibrium behavior, we address the practical concern of performance comparison between digital representatives and direct participation, leading to our third research question:

\begin{kq}
	How well do digital representatives perform in collective decision-making compared to direct member participation? 
\end{kq}

We analytically and numerically compare team decisions, preference losses, and total losses (preference loss plus communication cost) under two scenarios: Digital representation versus direct member participation.
Our results reveal several important insights with practical implications. 
First, digital representatives can closely approximate manual team decisions, particularly when members' preferences conflict. 
Second, although team preference loss is generally larger with digital representatives, the overall team loss can be lower when manual participation costs are high, members' preferences align with GenAI's prior, or when GenAI systems achieve sufficient capability. 
These findings provide critical guidance for deploying GenAI in collaborative decision environments and establish theoretical foundations for designing such representative systems.

This paper advances the state of the art in the following ways:
\begin{itemize}
	\item \textbf{New Theoretical Framework for Digital Representatives in Collective Decision-Making}: 
	To the best of our knowledge, this is the first theoretical investigation of strategic interactions between team members and GenAI-powered digital representatives in collective decision-making.
	We develop an information-centric model that quantifies representation fidelity as a function of revealed information, establishing a rigorous foundation for analyzing digital representation in collective settings. 
	This framework can be extended to address related challenges, including privacy preservation and heterogeneous member capabilities.	
	\item \textbf{Closed-Form Equilibrium Characterization with Strategic Insights}: 
	We derive analytical characterizations for members' optimal information revelation strategies at equilibrium.
	Our analysis reveals a counter-intuitive competitive dynamic: Members with aligned preferences strategically withhold information, benefiting from others' revelations without incurring costs themselves. In contrast, members with conflicting preferences engage in more intensive information revelation to prevent adverse decision outcomes. 
	This competitive revelation dynamic represents a previously unidentified strategic behavior in human-AI collaborative systems.
	\item \textbf{Comprehensive Performance Analysis with Practical Implications}: 
	We analytically and numerically compare digital representation against direct human participation across multiple performance dimensions: Team decision, preference loss, and total loss (including communication costs).
	Our results demonstrate that digital representatives can closely approximate manual team decisions when member preferences conflict, reducing overall team loss for time-intensive tasks or when GenAI systems are highly capable.
	These findings provide actionable guidance on when and how to deploy digital representatives in collective decision-making.

\end{itemize}

The remainder of this paper is organized as follows. 
We review the prior related research in Section \ref{sec: related work}.
In Section \ref{sec: model}, we formulate the model for GenAI as digital representatives. 
In Section \ref{sec: analysis}, we analyze members' equilibrium decisions, and in Section \ref{sec: performance}, we examine the outcomes of decision-making with or without GenAI. 
Section \ref{sec: simulation} provides numerical simulations. 
Finally, Section \ref{sec: conclusion} concludes the paper.


\section{Related Work}	\label{sec: related work}

This section contextualizes our work by reviewing prior research in three key areas: AI-powered collective decision-making, AI as a human proxy, and proxy voting.

\paragraph{AI-Powered Collective Decision-Making}

Recent studies have employed AI to support collective decision-making, often casting AI as an assistant (e.g., \cite{jannach2025rethinking, maragheh2025future, delic2023charm, bakker2022fine, burton2024large, tessler2024ai, jarrett2023language, papachristou2023leveraging, de2025supernotes}). 
For instance, AI has been implemented as a chat agent to guide group interactions \cite{delic2023charm}, as a facilitator to manage conversations \cite{papachristou2023leveraging}, or as a synthesis tool to generate consensus-building notes \cite{de2025supernotes}. 
This line of research primarily relies on experimental validation. 
\textit{In contrast, our work models AI as a proactive representative and focuses on the theoretical analysis of members' strategic decisions.}

\paragraph{GenAI as a Proxy for Humans}

A growing body of research has used GenAI to proxy human decision-makers in domains such as economic experiments (e.g., \cite{horton2023large, li2024econagent}), recommender systems (e.g., \cite{zhang2024generative, zhang2024usimagent}), and online advertising auctions (e.g., \cite{duetting2024mechanism, soumalias2024truthful}). 
These studies typically focus on the experimental implementation of GenAI proxies, the alignment between agents and their human counterparts, or high-level system designs where AI simulates human archetypes. 
\textit{Our model diverges by theoretically investigating how human users strategically reveal their preferences to their GenAI proxies.}

\paragraph{Proxy Voting}

Proxy voting is a collective decision-making paradigm where voters delegate their rights to a representative. 
Existing research—from game-theoretical \cite{bloembergen2019rational, zhang2021power}, topological (e.g., \cite{anshelevich2021representative, berinsky2025tracking}), or computational (e.g., \cite{kahng2021liquid, tyrovolas2024unravelling}) viewpoints—has focused mainly on the rationality of delegation for objective tasks \cite{bloembergen2019rational} or the optimal selection of proxies from a set of candidates \cite{anshelevich2021representative}. 
\textit{Our work, in contrast, focuses on personalizing a representative for subjective decisions that lack a ground truth, rather than simply selecting a pre-existing proxy.}


\section{System Model}	\label{sec: model}

We begin with an illustrative example to motivate the formulation.
Consider Alice and Bob planning a weekend activity.
Alice prefers introverted activities (e.g., 80\% introvert), while Bob prefers extroverted ones (e.g., 70\% extrovert).\footnote{
	These quantitative numbers can be understood as the numerical values of the Myers-Briggs Type Indicator test results, which are rooted in the personality of each member.
	Other preferences (e.g., film tastes) are also encoded in personal information (e.g., past viewing records and ratings).}
Each of them deploys a GenAI digital representative and strategically decides how much personal information (e.g., activity histories) to share with the GenAI.
The digital representatives then learn from the revealed information and express inferred preferences, which are aggregated into a final team decision.

We now formalize this process within a rigorous mathematical framework, as illustrated in Figure \ref{fig: 2-model}.

\paragraph{Member Preferences}
Alice (A) and Bob (B), with preferences $\theta_A, \theta_B \in \mathbb{R}$,  respectively, collaborate on a joint decision (e.g., entertainment choices).\footnote{We refer to Alice as "she" and Bob as "he" for clarity.
}
While real-world preferences are high-dimensional as an initial step, we follow prior work \cite{estornell2024multi, jiang2023latent, xie2022an} and model each user's preference as a single latent feature $\theta$ representing core inclinations such as social energy (in our motivating example), political orientation, or aesthetic taste.

\paragraph{GenAI as Digital Representatives}
Alice and Bob utilize GenAI as digital representatives to aid decision-making by revealing their preferences.
Inspired by \cite{castro2024human}, we model the interaction between a member and GenAI as follows.
GenAI starts with a prior understanding $\mu_G$ about general population preferences for the task \cite{10.1145/3670865.3673532}.
When member $m\in \mathcal{M} \triangleq \{A, B\}$ with preference $\theta_m$ decides on the revelation level $\alpha_m \in \mathcal{S}_m \triangleq [0,1]$ to GenAI, the digital representative learns an updated preference $\theta_m'$ whose expected value is a weighted average between the member's true preference $\theta_m$ and the prior  $\mu_G$ \cite{berger2013statistical}:

\begin{equation}	\label{eq: dr output}
	E[\theta_m'] = \lambda(\alpha_m) \theta_m + [1-\lambda(\alpha_m)]\mu_G,
\end{equation}
where the weight function $\lambda(\alpha_m)$ is assumed to increase concavely with the revelation level.
In more detail, if a member fully reveals information ($\alpha_m = 1$), the digital representative expresses the exact preference ($E[\theta_m'] = \theta_m$), while no revelation ($\alpha_m = 0$) results in the output of the prior ($E[\theta_m'] = \mu_G$).

As is common in the literature (e.g., \cite{castro2024human, gentzkow2014costly}), we formulate the communication cost from an information exchange perspective.
We assume the communication cost $c(\alpha_m)$ increases convexly with the revelation level.\footnote{For simplicity, we assume Alice and Bob face identical weight $\lambda(\cdot)$ and cost $c(\cdot)$ functions, corresponding to using the same GenAI system with similar interaction abilities. 
	Extensions to heterogeneous cases would retain the core insights while introducing additional complexity.
}
Full revelation ($\alpha_m = 1$) incurs infinite communication cost ($c(1) = +\infty$), while no revelation ($\alpha_m = 0$) requires no cost ($c(0) = 0$).

\paragraph{Team Decision Formation}
The final team decision $\theta_T$ is the average of the digital representatives' expected outputs:\footnote{We focus on subjective tasks with equal member importance, averaging the representatives' outputs. 
	GenAI may implement this aggregation as in \cite{bakker2022fine}.
	Models using other aggregation rules deserve further exploration.}
\begin{equation}	\label{eq: dr decision}
	\theta_T(\bm{\alpha}) = \frac{E[\theta_A'] + E[\theta_B']}{2},
\end{equation}
where $\bm{\alpha} \triangleq (\alpha_m, \forall m \in \mathcal{M})$ denotes the revelation profile.

\paragraph{Member's Loss}
Each member's \textit{total loss} $L_m(\bm{\alpha})$ includes the \textit{preference loss} $e_m(\bm{\alpha}) \triangleq \left[ \theta_m - \theta_T(\bm{\alpha}) \right]^2$ and the \textit{communication cost} $c(\alpha_m)$:
\begin{equation}	\label{eq: loss}
	L_m(\bm{\alpha}) =  \left[ \theta_m - \theta_T(\bm{\alpha}) \right]^2  + c(\alpha_m).
\end{equation}

As Alice and Bob's revelation decisions affect both of their losses, we formalize their interactions as a digital representative game:

\begin{definition}[Digital Representative Game]	\label{def: dr game}
	The Digital Representative Game $\Gamma(\mathcal{M}, \mathcal{S}, \bm{L})$ is defined by:
	\begin{itemize}
		\item Players: The set $\mathcal{M}$ of members.
		\item Strategies: Each member $m\in \mathcal{M}$ chooses the revelation level $\alpha_m\in \mathcal{S}_m$ to GenAI.
		The feasible set of all strategy profiles is $\mathcal{S} = \prod_{m\in \mathcal{M}} \mathcal{S}_m$.
		\item Payoffs: The vector $\boldsymbol{L} = (L_m,\ \forall m \in \mathcal{M})$ contains each member's total loss as defined in Eq. \eqref{eq: loss}.
	\end{itemize}
\end{definition}

In the next section, we analyze Alice and Bob's equilibrium strategies and derive insights into the implications of using GenAI as a digital representative in collective decision-making.


\section{Game-Theoretical Analysis}	\label{sec: analysis}
To streamline the presentation, we define several key terms: 
\textit{1)} The \textit{trigger price} $p \triangleq \frac{c' (0)}{\lambda' (0)}$, which represents the ratio between the initial cost of revealing a preference and the benefit of aligning the GenAI's decision with a member's preference.
\textit{2)} Member $ m$'s \textit{directional diversity} $d_m \triangleq \theta_m - \mu_G$, which is the difference between a member's preference and the GenAI's prior;
\textit{3)} \textit{The diversity} $|d_m|$ measuring the absolute difference between a member's preference and the GenAI's prior;
and \textit{4)} The \textit{preference deviation ratio} $\kappa_{m, -m} = \frac{d_m}{d_{-m}}$, which captures the relative directional diversity of the two members.\footnote{When a member's preference aligns exactly with the GenAI's prior ($d_m=0$), the optimal strategy is trivial: No information revelation is necessary. 
	The subsequent analysis addresses the case where $d_m\neq 0$.
}

We first analyze the best response in Section~\ref{subsec: br analysis}, and then derive the Nash equilibrium of the digital representative game in Section~\ref{subsec: ne}.

\subsection{Best Response Analysis}	\label{subsec: br analysis}
In this section, we focus on analyzing Alice's best response, noting that Bob's response follows symmetrically. We formally define:

\begin{definition}[Alice's Best Response]	\label{def: br}
	Given Bob's revelation decision $\alpha_B$, Alice's best response $\alpha_A^*(\alpha_B)$ is 
	\begin{equation}	\label{eq: br a}
		\alpha_A^*(\alpha_B) = \arg\min_{\alpha_A\in [0,1]} L_A(\alpha_A, \alpha_B).
	\end{equation}
\end{definition}

To facilitate analysis, we introduce: 

\begin{itemize}
	\item Revelation threshold $\tilde{\alpha}_m \triangleq \lambda^{-1}(\frac{2}{\kappa_{m, -m}} - \frac{2p}{d_A d_B})$: The threshold determining when Member m starts or stops revealing information.
	\item Function $f_m(\alpha_m) \triangleq \lambda(\alpha_m) + \frac{2}{d_m^2} \frac{c'(\alpha_m)}{\lambda'(\alpha_m)}$: Member $m$'s benefit from revealing information.
	\item Function $h_m(\alpha_{-m}) \triangleq 2- \frac{\lambda(\alpha_{-m})}{ \kappa_{m, -m}}$: The strategic impact of the other member's revelation on Member $m$.
\end{itemize}

By solving Eq.~\eqref{eq: br a}, we derive Alice's best response in Proposition~\ref{prop: br}.

\begin{proposition}[Best Response of Digital Representative Game]	\label{prop: br}
	Let $\alpha_A^*(\alpha_B) \triangleq f_A^{-1} ({h}_A(\alpha_B))$ denote "\underline{R}evealing information strategy of \underline{A}lice (R-A)".
	Given Bob's revelation decision $\alpha_B$, Alice's best response $\alpha_A^*(\alpha_B)$ is:
	\begin{enumerate}
		\item If Alice and Bob have conflicting preferences, $\kappa_{A, B} \in (-\infty, 0)$,
		\begin{equation}	\label{eq: op br}
			\alpha_A^*(\alpha_B) = 
			\begin{cases}
				0, & \text{if } d_A^2(1 - \frac{1}{2\kappa_{A, B}}) \le p, \\
				& \text{or } d_A^2 < p < d_A^2(1 - \frac{1}{2\kappa_{A, B}}) \text{ and } \alpha_B \le \tilde{\alpha}_B,\\
				\text{R-A}, & \text{if } d_A^2 \ge p, \\
				& \text{or } d_A^2 < p < d_A^2(1 - \frac{1}{2\kappa_{A, B}}) 
				\text{ and } \alpha_B > \tilde{\alpha}_B.\\
			\end{cases}
		\end{equation}
		
		\item If Alice and Bob have aligned preferences, $\kappa_{A, B} \in (0, +\infty)$,
		\begin{equation}	\label{eq: ss br}
			\alpha_A^*(\alpha_B) = 
			\begin{cases}
				0, & \text{if } d_A^2 \le p, \\
				& \text{or } d_A^2(1 - \frac{1}{2\kappa_{A, B}}) < p < d_A^2 
				\text{ and } \alpha_B > \tilde{\alpha}_B,\\
				\text{R-A}, & \text{if } d_A^2(1 - \frac{1}{2\kappa_{A, B}}) \ge p,\\
				& \text{or } d_A^2(1 - \frac{1}{2\kappa_{A, B}}) < p < d_A^2 
				\text{ and } \alpha_B \le \tilde{\alpha}_B.
			\end{cases}
		\end{equation}
		
	\end{enumerate}
\end{proposition}

The proof is provided in Section~A of the technical report \cite{techrepo}.

Our analysis reveals several important findings. 
First, we demonstrate that when Alice's diversity ($|d_A|$) is small, she opts not to reveal information, minimizing communication costs, regardless of Bob's strategy. 
Alice partially reveals her preferences as her diversity increases, balancing representation with costs.\footnote{For fixed parameters $d_A$ and $d_B$, $f_A(1)$ reaches infinity while $h_A(\alpha_B)$ is bounded.
	Thus, we can prove that $\alpha_A^*<1$, i.e., Alice only partially reveals her information.}

Second, we uncover a more nuanced insight when comparing conflicting versus aligned preferences cases. 
When preferences conflict ($\kappa_{A, B} < 0$), Alice begins to reveal information if Bob's revelation exceeds a threshold $\tilde{\alpha}_B$. 
In contrast, when preferences are aligned ($\kappa_{A, B} > 0$), Alice initiates revelation when Bob's revelation falls below a threshold.
This reveals a key finding: Conflicting preferences lead to competitive behavior—high revelation by Bob pulls the team decision away from Alice's preference. In contrast, aligned preferences promote free-riding, with Alice relying on Bob's revelation unless it is insufficient.

With this foundational decision logic in mind, we proceed to analyze the Nash equilibrium of the digital representative game.

\subsection{Nash Equilibrium}	\label{subsec: ne}
Due to space limitations, we focus on cases where $\kappa_{A, B} \in (-\infty, -1] \cup [1,+\infty)$, i.e., Alice has a more diverse preference than Bob.
Equilibrium results for other $\kappa_{A, B}$ ranges are analogous.\footnote{For $\kappa_{A, B}\in (-1, 0) \cup (0,1)$, where Bob's preference is more diverse, the equilibrium results are symmetric.
	When $d_m = 0$ for any member, non-revelation is optimal.}
Detailed derivations are provided in Section~B.1 of the technical report \cite{techrepo}.

\begin{definition}[Nash Equilibrium]	\label{def: pne}
	An information revelation profile $\boldsymbol{\alpha}^* = (\alpha_m^*,\alpha_{-m}^*)$ constitutes a Nash Equilibrium (NE), if for each member $m \in \mathcal\{A, B\}$ and revelation decision $\alpha_m \in [0,1]$, 
	\begin{equation}
		L_m(\alpha_m^*,\alpha_{-m}^*) \le L_m(\alpha_m',\alpha_{-m}^*),
	\end{equation}	
	where $\alpha_{-m}^*$ is the equilibrium strategy of the other member.
\end{definition}

\paragraph{Equilibrium Characterization}
Intuitively, the fixed point of both members' best response choices is
the Nash equilibrium, constituting a stable strategy profile where no member can reduce their loss by unilaterally changing their strategy.

Using the best responses from Proposition~\ref{prop: br}, we characterize the Nash equilibria in Theorem~\ref{thm: 2-ne}.

\begin{theorem}[Nash Equilibrium of Digital Representative Game]	\label{thm: 2-ne}
	We define three equilibrium profiles:
	\begin{itemize}
		\item "\underline{N}o member \underline{R}eveals information (NR)" equilibrium: $\bm{\alpha}^* = (0, 0)$.
		\item "\underline{O}ne member \underline{P}artially \underline{R}eveals information (OPR)" equilibrium: $\bm{\alpha}^* = (f_{A}^{-1}(2), 0)$.\footnote{The equilibrium profile $\bm{\alpha}^* = (0, f_{B}^{-1}(2))$, which happens when Bob's preference is more diverse, i.e., $(-1, 0) \cup (0,1)$, also belongs to the OPR equilibrium.}
		\item "\underline{B}oth member \underline{P}artially \underline{R}eveal information (BPR)" equilibrium: $\bm{\alpha}^* = (f_A^{-1}({h}_A(\alpha_B^*)),f_{B}^{-1}({h}_{B}(\alpha_A^*)))$.
	\end{itemize} 
	
	The Nash equilibrium of the Digital Representative Game is:
	
	\begin{enumerate}
		\item For $\kappa_{A, B} \in (-\infty, -1]$, the equilibrium results are:
		\begin{equation}	\label{eq: ss ne}
			\bm{\alpha}^* = 
			\begin{cases}
				\text{NR}, & \text{if } d_A^2 < p\ \text{and }  d_{B}^2 < p,\\
				\text{OPR}, & \text{if }d_A^2 \ge p\ \text{and } d_{B}^2(1 - \frac{\kappa_{A, B}}{2}) \le p, \\
				& \text{or } d_A^2 \ge p,\ d_{B}^2 < p < d_{B}^2(1 - \frac{\kappa_{A, B}}{2}), \\
				& \text{and } f_A^{-1}(2) \le \tilde{\alpha}_A, \\
				\text{BPR}, & \text{if } d_A^2 \ge p\ \text{and } d_{B}^2 \ge p,\\
				& \text{or } d_A^2 \ge p,\ d_{B}^2 < p < d_{B}^2(1 - \frac{\kappa_{A, B}}{2}), \\
				& \text{and } f_A^{-1}(2) > \tilde{\alpha}_A.
			\end{cases}
		\end{equation}

		\item For $\kappa_{A, B} \in [1, +\infty)$, the equilibrium results are:
		\begin{equation}	\label{eq: op ne}
			\bm{\alpha}^* = 
			\begin{cases}
				\text{NR}, & \text{if } d_A^2 \le p\ \text{and } d_{B}^2 \le p,\\
				\text{OPR}, & \text{if } d_A^2 > p\ \text{and } d_{B}^2 \le p,\\
				& \text{or } d_A^2 > p,\ d_{B}^2 (1 - \frac{\kappa_{A, B}}{2}) < p < d_{B}^2,\\
				& \text{and } f_{A}^{-1}(2) \ge \tilde{\alpha}_A,\\
				\text{BPR}, & \text{if } d_A^2(1 - \frac{1}{2\kappa_{A, B}}) \ge p\ \text{and } d_{B}^2(1 - \frac{\kappa_{A, B}}{2}) \ge p,\\
				& \text{or } d_A^2 > p,\ d_{B}^2 (1 - \frac{\kappa_{A, B}}{2}) < p < d_{B}^2,\\
				& \text{and } f_{A}^{-1}(2) < \tilde{\alpha}_A.\\
			\end{cases}
		\end{equation}
		
	\end{enumerate}
	
\end{theorem}

\begin{figure}[t]
	\centering
	\includegraphics[width=.8\linewidth]{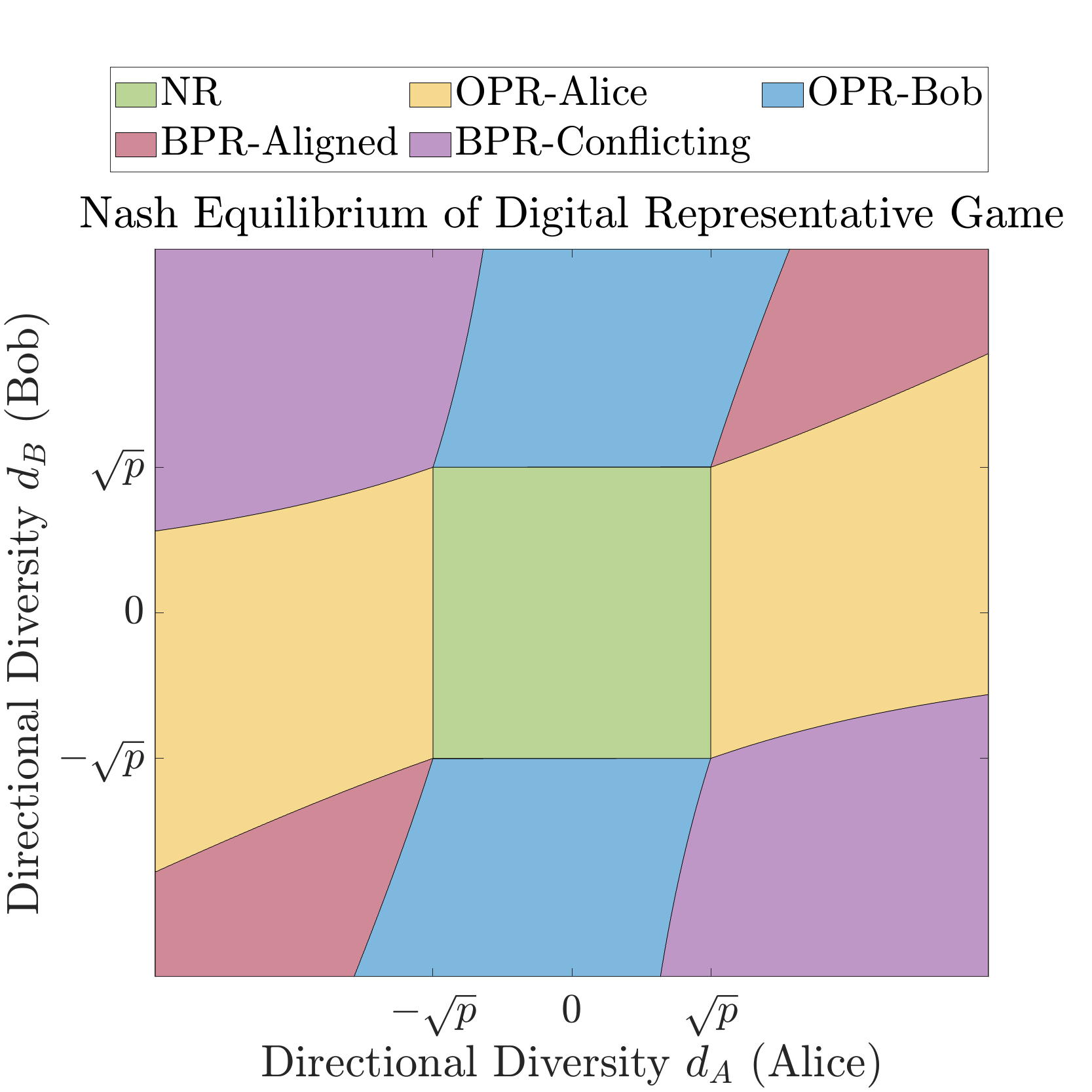}
	\caption{Illustration of two-member Nash equilibrium: 
		No revelation (green), one-member revelation (yellow for Alice, blue for Bob), and both members revealing with aligned (red) or conflicting (purple) preferences.
	}
	
	\label{fig: 2-ne}
\end{figure}

The complete proof is provided in Section~B.1 of the technical report \cite{techrepo}.
Figure \ref{fig: 2-ne} illustrates the equilibrium results visually.

From Theorem \ref{thm: 2-ne}, we observe that the equilibrium outcomes are sensitive to the trigger price $p$, which is influenced by many practical factors, such as GenAI capability, members' proficiency in utilizing GenAI, and members' privacy concerns.
For instance, when the GenAI is highly user-friendly (e.g., featuring voice-based interaction) or demonstrates strong learning capabilities, the trigger price becomes lower compared to less capable systems.
Consequently, members are more willing to reveal information.
Conversely, when the barrier to digital representative personalization is high (i.e., a higher trigger price), members exhibit greater reluctance to reveal personal information.

\paragraph{Equilibrium Properties}
Due to the general forms assumed for the weight function $\lambda(\cdot)$ and cost function $c(\cdot)$, obtaining closed-form expressions for the exact equilibrium levels of OPR and BPR is analytically challenging.
We present the following propositions to characterize equilibrium properties to address this challenge and enhance understanding.

\begin{proposition}[OPR Equilibrium Properties]	\label{prop: opr trend}
	When only one member reveals information at equilibrium, the revelation level increases with the member's diversity.
\end{proposition}

The proof is provided in Section~B.2.1 of the technical report \cite{techrepo}.

OPR equilibrium occurs when one member's diversity exceeds a threshold, while the other's is relatively moderate (yellow/blue regions in Figure \ref{fig: 2-ne}).
This equilibrium has an important interpretation (as shown by the yellow regions): 
As Alice's diversity grows, the GenAI becomes less representative of her preferences. 
Consequently, she reveals more information to reduce the preference loss $e(\bm{\alpha})$.

\begin{proposition}[BPR Equilibrium Properties]	\label{prop: bpr trend}
	When both members reveal information at equilibrium:
	\begin{enumerate}
		\item With aligned preferences, each member's equilibrium revelation increases with their own diversity but decreases with the other's diversity.
		\item With conflicting preferences, each member's equilibrium revelation increases with both members' diversity.
		\item For equal $|\kappa_{A, B}|$, conflicting members reveal more information than those with aligned preferences.
	\end{enumerate}
\end{proposition}

The proof is provided in Section~B.2.2 of the technical report \cite{techrepo}.

BPR equilibrium emerges when both members' preferences significantly deviate from the GenAI's prior (red/purple regions in Figure~\ref{fig: 2-ne}).
With aligned preferences (red region), members can leverage each other's revelations, so a member strategically reduces revelation when the other reveals more.
In contrast, with conflicting preferences (purple region), members engage in more competitive behavior, strategically increasing information revelation to prevent the team decision from diverging from their preferences.

\begin{proposition}[Diversity and Revelation]	\label{prop: two member comp}
	If Member $m$'s preference is more diverse ($|d_m| \ge |d_{-m}|$), the equilibrium revelations satisfy:
	\begin{equation}	\label{eq: am>a-m}
		\alpha_m^* \ge \alpha_{-m}^*.
	\end{equation}
\end{proposition}

The proof is provided in Section~B.2.3 of the technical report \cite{techrepo}.

This proposition offers an intuitive interpretation:
Since the default GenAI system deviates more from the preference of a member with greater diversity (e.g., Alice) compared to one with less diversity (e.g., Bob), Alice is incentivized to reveal more information for more representative outputs.

Our theoretical analysis culminates in a significant finding from Theorem \ref{thm: 2-ne}: No member fully discloses their private information at equilibrium, resulting in imperfect representation by the GenAI.
This discrepancy highlights the importance of understanding members' information revelation behaviors.
Identifying and addressing these gaps can guide better designs of future collective digital systems.

Based on the equilibrium results, we proceed to address our key question regarding the digital representatives' equilibrium outcomes.


\section{Equilibrium Outcomes of Digital Representatives}	\label{sec: performance}
This section comprehensively evaluates equilibrium performance with and without GenAI as digital representatives to assess their viability for future applications. 
Section \ref{subsec: baseline} establishes a theoretical benchmark by examining decision-making without GenAI, while Section \ref{subsec: comp} systematically compares equilibrium outcomes to extract actionable insights for implementing digital representatives in collective decision-making scenarios.

\subsection{Baseline}	\label{subsec: baseline}

We establish the scenario without GenAI as the baseline, where each member directly and completely expresses their preferences. 
However, manual decision-making necessitates a constant cost, denoted by $C$, which depends on the task but is considered identical for both members \cite{castro2024human}.

To facilitate our analysis, we define the team preference loss as $e_T\triangleq e_A + e_B$ and the team total loss $L_T \triangleq L_A + L_B$.
Results in the baseline setting are distinguished with superscript "$B$" and are given by:
\begin{equation}	\label{eq: baseline}
	\begin{cases}
		\theta_T^B = \frac{\theta_A + \theta_B}{2},\\
		e_T^B = \frac{(\theta_B - \theta_A)^2}{2},\\
		L_T^B = \frac{(\theta_B - \theta_A)^2}{2} + 2C.
	\end{cases}
\end{equation}

With equal preference aggregation, the baseline team decision $\theta_T^B$ achieves the optimal team preference loss $e_T$.
As shown in Eq.~\eqref{eq: baseline}, both team preference loss and team total loss are directly proportional to the distance between Alice's and Bob's preferences.

Having established this benchmark, we compare the equilibrium and baseline team outcomes.

\subsection{Outcome Comparison}	\label{subsec: comp}

For simplicity, let $\lambda(\alpha_m^*) \triangleq \lambda_m^*$ and $c(\alpha_m^*) \triangleq c_m^*$ denote the equilibrium weight and cost for member $m\in\{A, B\}$.

\begin{theorem}[Equilibrium versus Baseline Team Outcomes]	\label{thm: comp}
	The comparisons between the equilibrium and baseline team outcomes are summarized as follows:
	\begin{enumerate}
		\item The gap between team decisions is:
		\begin{equation}	\label{eq: thetat comp}
			|\theta_T^* - \theta_T^B| = |\frac{(1-\lambda_A^*)d_A + (1-\lambda_{B}^*)d_{B}}{2}|.
		\end{equation}
		
		\item The team preference loss satisfies:
		\begin{equation}	\label{eq: et comp}
			e_T^* \ge e_T^B, 
		\end{equation}
		with equality when $\kappa_{A, B} = -1$ in NR equilibrium, $\kappa_{A, B} = -\frac{1}{1-\lambda_A^*}$ in OPR-A equilibrium, $\kappa_{A, B} = -(1-\lambda_B^*)$ in OPR-B equilibrium, or $\kappa_{A, B} = -\frac{1-\lambda_B^*}{1-\lambda_A^*}$ in BPR-Conflicting equilibrium.
		
		\item The team total loss satisfies:
		\begin{equation}	\label{eq: lt comp}
			\begin{cases}
				L_T^* \le L_T^B, \quad & \text{if } C \ge \frac{[(1-\lambda_A^*) d_A + (1-\lambda_{B}^*) d_{B}]^2}{4} + \frac{c_A^* + c_B^*}{2} ,\\
				L_T^* > L_T^B, \quad & \text{otherwise}.
			\end{cases}
		\end{equation}
	\end{enumerate}
	
\end{theorem}

Detailed derivations are provided in Section~C of the technical report \cite{techrepo}. 
Below, we highlight the key insights from Theorem~\ref{thm: comp}.

First, from Theorem~\ref{thm: 2-ne}, no member fully reveals their preferences at equilibrium ($\alpha_m^*<1$). 
Consequently, digital representatives cannot perfectly reflect members' preferences, resulting in a gap between team decisions, as described in Eq.~\eqref{eq: thetat comp}.

Second, the baseline team preference loss $e_T^B$ represents the theoretical minimum under equal preference aggregation. 
Thus, the equilibrium preference loss $e_T^*$ is generally higher, as shown in Eq.~\eqref{eq: et comp}, except in specific conflicting scenarios where equality holds.
Most notably, in BPR-Aligned equilibrium, the team preference loss is strictly greater than in the baseline due to members' partial revelation, contradicting the intuition that aligned preferences should lead to better outcomes.

Finally, Eq.~\eqref{eq: lt comp} yields a critical threshold condition that digital representatives yield lower total team loss when the manual decision cost $C$ is sufficiently high.
High-cost tasks, such as strategic planning or complex negotiations, necessitate extensive effort, making digital representatives superior.
Conversely, low-cost tasks, such as routine decisions, favor direct participation.\footnote{These examples are illustrative; actual costs vary depending on task complexity and team context. 
	While GenAI is often associated with automating simple tasks, this study focuses on its potential as a digital representative in collective decision-making.}

Thus far, we have developed a comprehensive theoretical framework for the digital representative game. 
Numerical simulations in the next section further validate and illustrate these insights.


\begin{figure*}[t]
	\begin{minipage}[t]{0.33\textwidth}
		\centering
		\includegraphics[width=\linewidth]{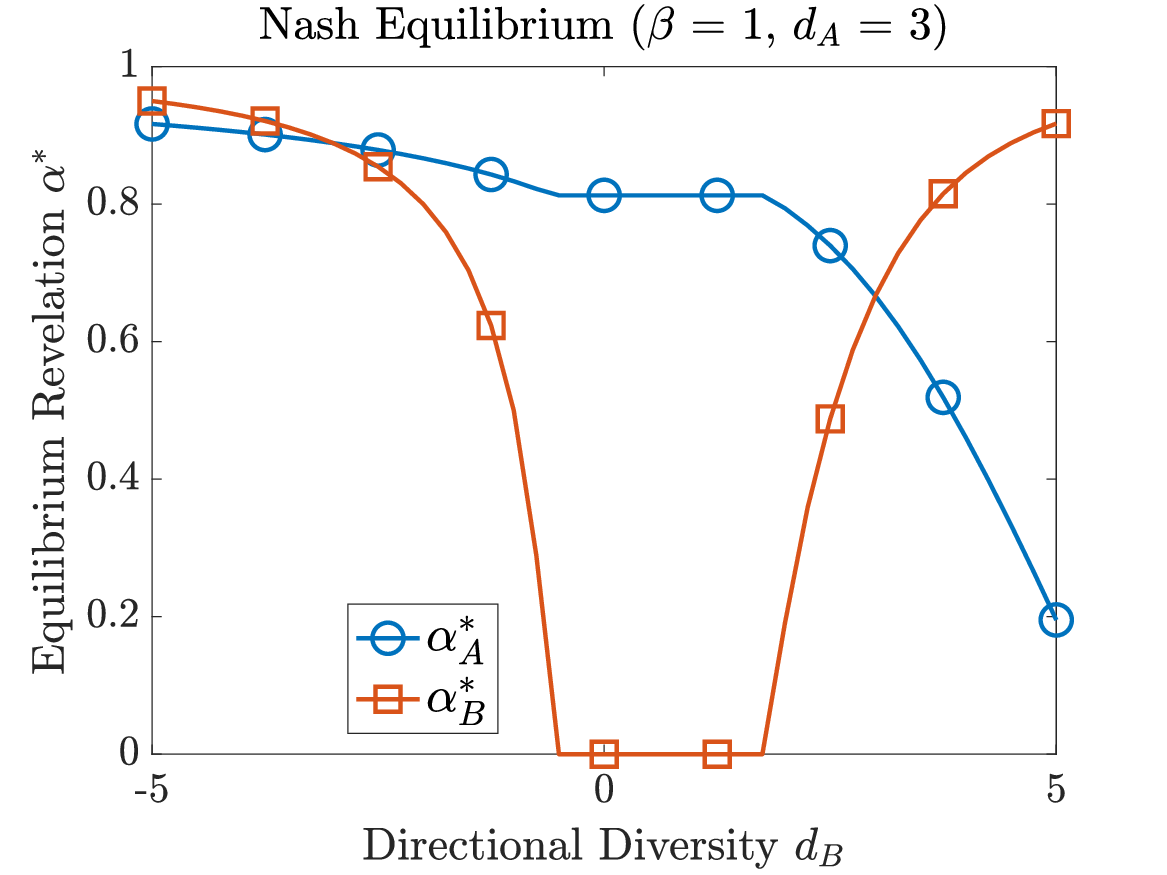}
		\caption{Illustration of Nash Equilibrium.}
		\label{fig: 2-ne-a}
	\end{minipage}
	\hfill
	\begin{minipage}[t]{0.33\textwidth}
		\centering
		\includegraphics[width=\linewidth]{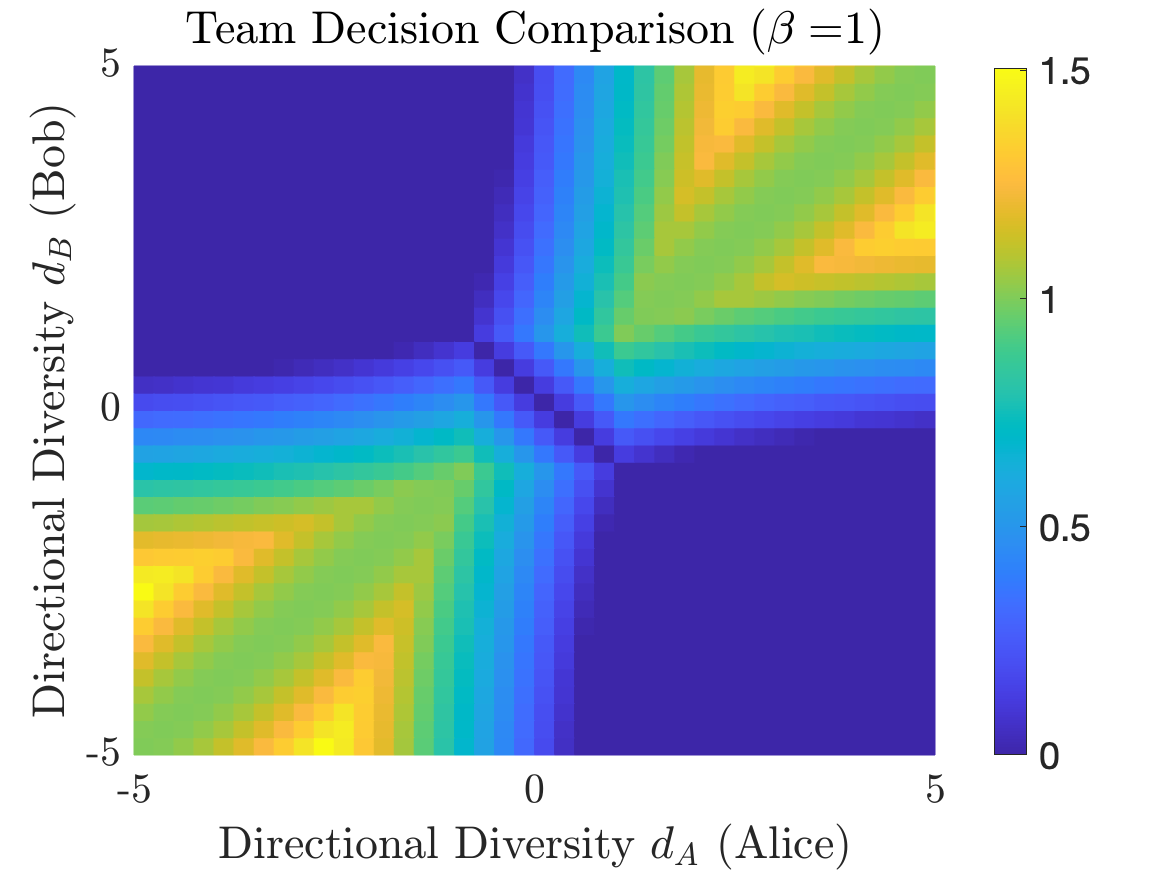}
		\caption{Comparison of Team Decision.}
		\label{fig: 2-ne-thetat-comp}
	\end{minipage}
	\hfill
	\begin{minipage}[t]{0.33\textwidth}
		\centering
		\includegraphics[width=\linewidth]{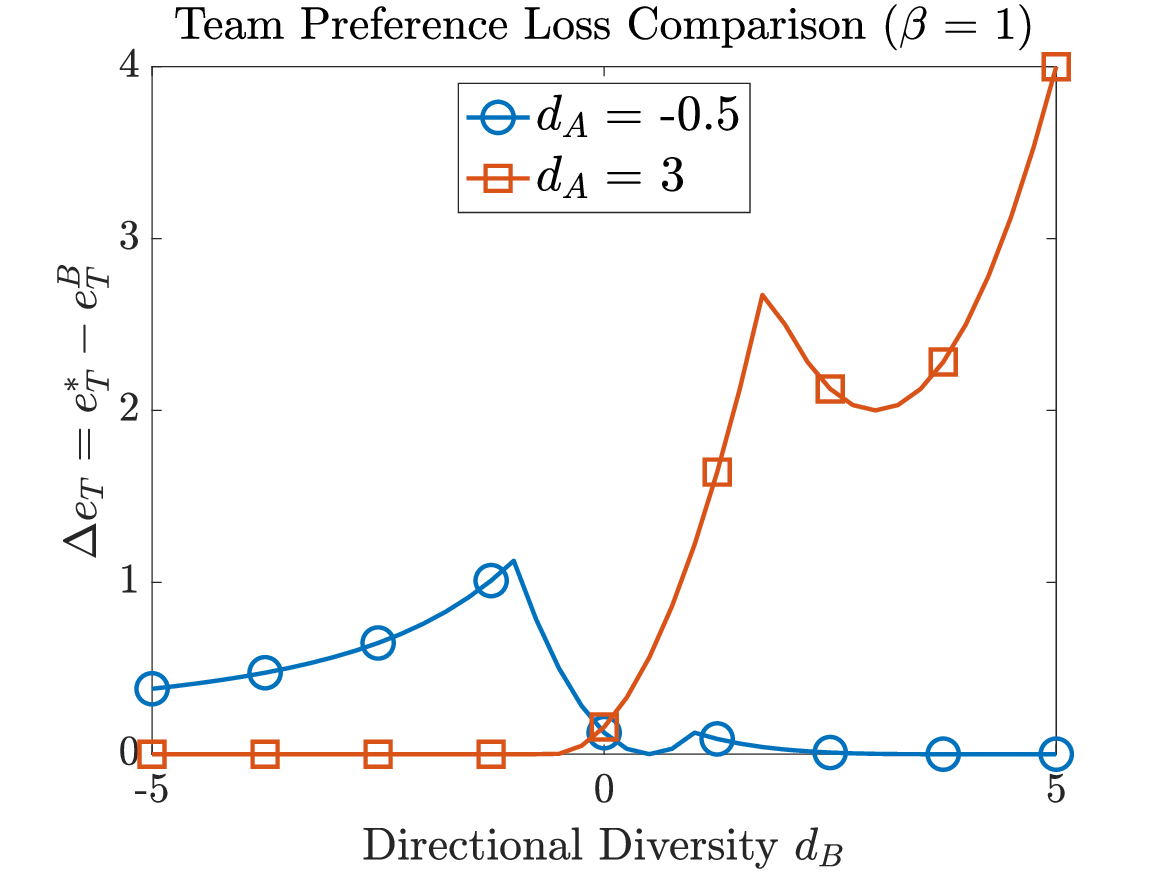}
		\caption{Comparison of Team Preference Loss.}
		\label{fig: 2-ne-et-comp}
	\end{minipage}
	
	\begin{minipage}[t]{0.33\textwidth}
		\centering
		\includegraphics[width=\linewidth]{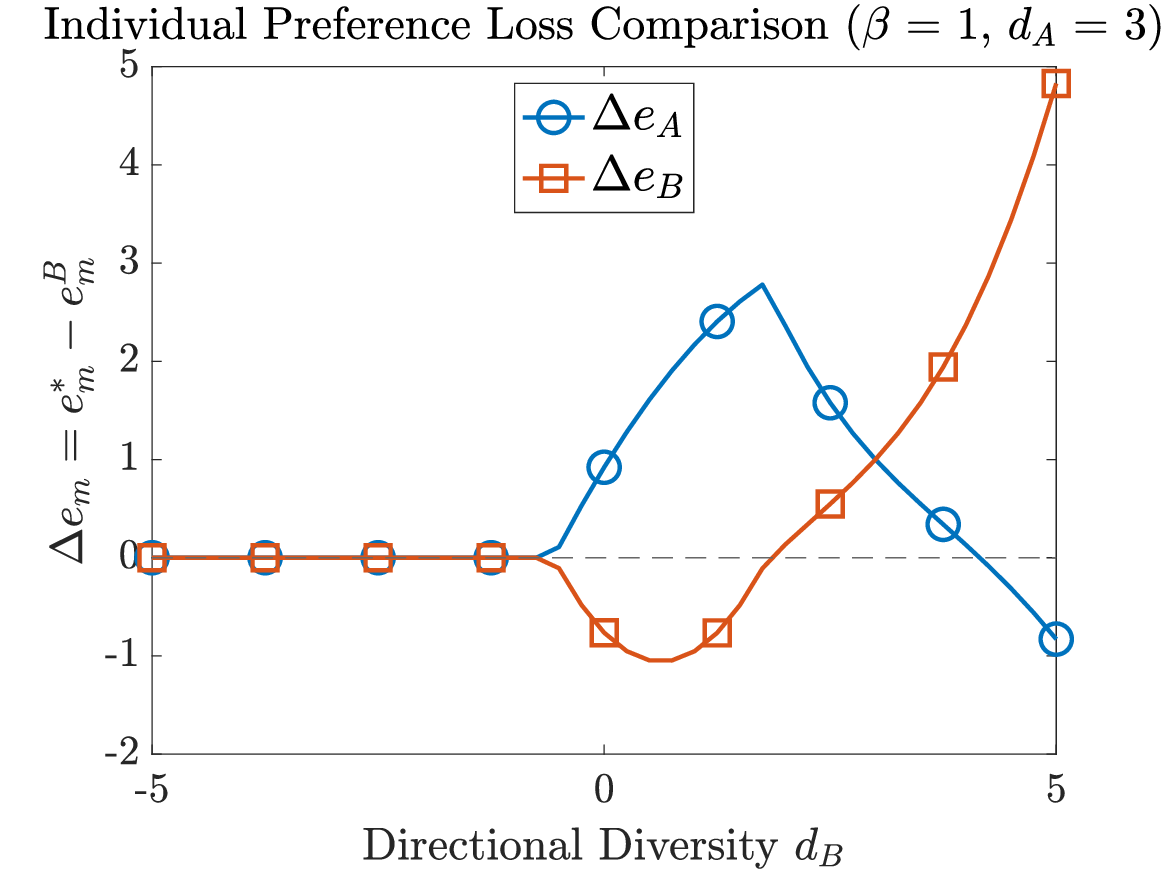}
		\caption{Comparison of Individual Preference Loss.}
		\label{fig: 2-ne-em-comp}
	\end{minipage}
	\hfill
	\begin{minipage}[t]{0.33\textwidth}
		\centering
		\includegraphics[width=\linewidth]{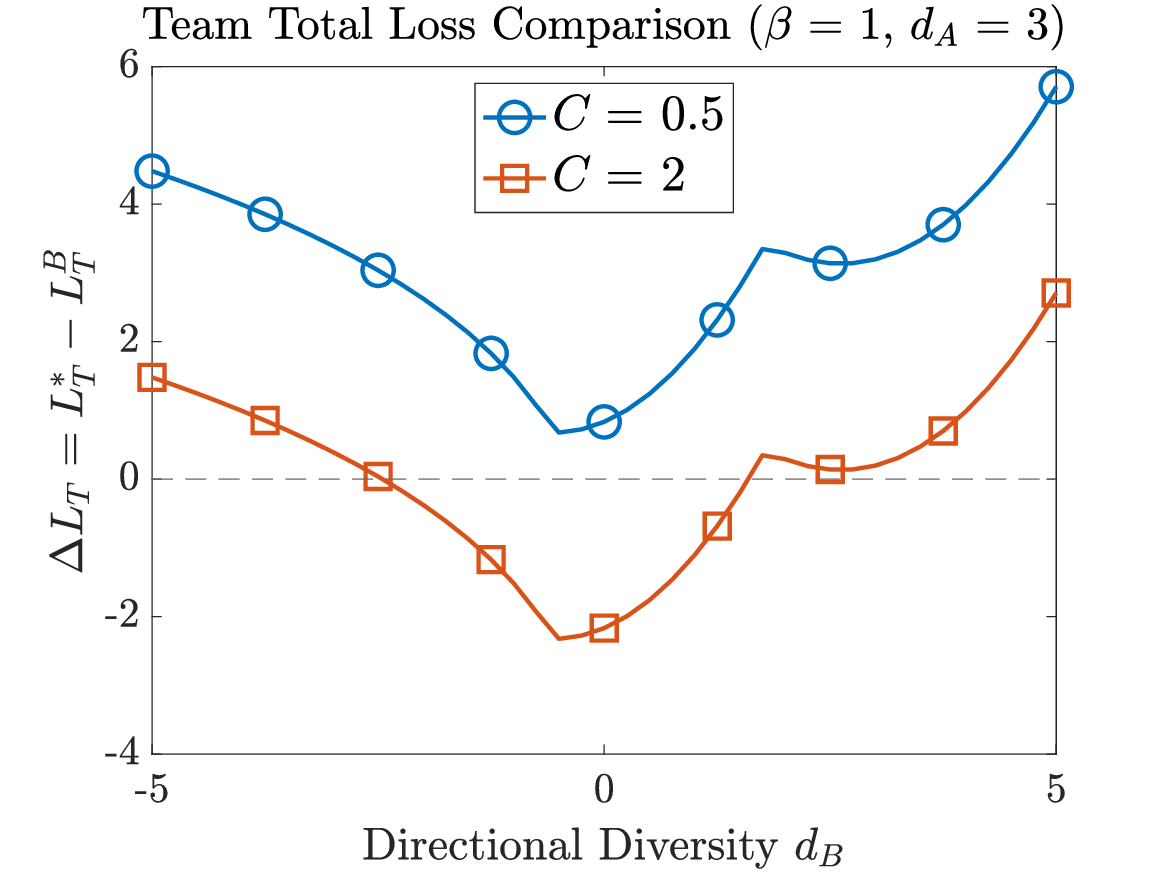}
		\caption{Comparison of Team Total Loss.}
		\label{fig: 2-ne-lt-comp}
	\end{minipage}
	\hfill
	\begin{minipage}[t]{0.33\textwidth}
		\centering
		\includegraphics[width=\linewidth]{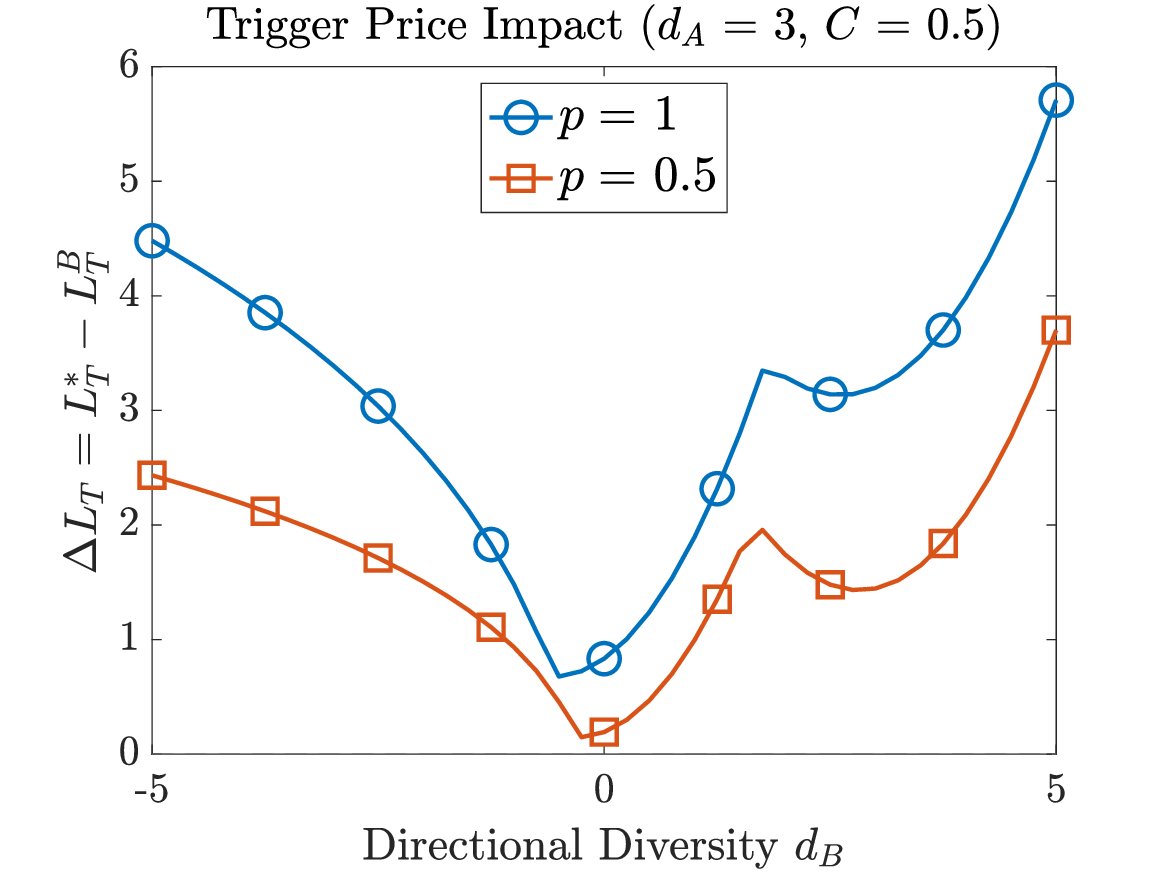}
		\caption{Comparison of Trigger Price.}
		\label{fig: 2-ne-beta-comp}
	\end{minipage}
\end{figure*}

\section{Numerical Results}	\label{sec: simulation}
In this section, we present numerical results to support our theoretical findings. 
For illustrative purposes, we adopt a linear weight function $\lambda(\alpha_m) = \alpha_m$ and a logarithmic cost function $c(\alpha_m) = -\beta \ln (1-\alpha_m)$ \cite{castro2024human, gentzkow2014costly}, where $\beta$ represents the cost of interacting with GenAI.\footnote{The parameter $\beta$ also serves as a tunable weight balancing members' preference loss and communication costs.}
The equilibrium results are provided in Section~\ref{subsec: num ne}, followed by a performance comparison in Section~\ref{subsec: num comp}.

\subsection{Equilibrium Results}	\label{subsec: num ne}
With $\beta = 1$ and fixing Alice's directional diversity at $d_A = 3$, we visualize the equilibrium revelation levels $\bm{\alpha}^*$ in Figure \ref{fig: 2-ne-a}.

\begin{obs}	\label{obs: 2-ne}
	From Figure \ref{fig: 2-ne-a}, we observe:
	\begin{enumerate}
		\item The member with greater preference diversity ($|d_m|$) consistently demonstrates higher information revelation at equilibrium, confirming our theoretical prediction about diversity-driven behavior.
		\item Bob's equilibrium revelation positively correlates with his diversity $|d_B|$, reinforcing the fundamental relationship between preference deviation and strategic information sharing.
		\item With aligned preferences ($d_B > 0$), Alice's revelation decreases as Bob's diversity increases, illustrating the strategic substitution effect where one member reduces revelation as the other increases.
		\item With conflicting preferences ($d_B < 0$), Alice's revelation counter-intuitively increases with Bob's diversity, highlighting the competitive dynamics driven by opposing interests.
	\end{enumerate}
\end{obs}

These findings corroborate the analytical results in Section~\ref{subsec: ne}.

\subsection{Outcome Comparison Results}	\label{subsec: num comp}
We compare equilibrium outcomes with the baseline, using $\beta = 1$, across team decision, preference loss, and team total loss metrics.

\paragraph{Team Decision Comparison}	\label{para: num thetat comp}
The gap between the equilibrium team decision $\theta_T^*$ and the baseline $\theta_T^B$ is shown in Figure \ref{fig: 2-ne-thetat-comp}.

\begin{obs}	\label{obs: thetat comp}
	From Figure \ref{fig: 2-ne-thetat-comp}, we observe: 
	\begin{enumerate}
		\item The equilibrium team decision perfectly aligns with the baseline exclusively in conflicting preference scenarios (dark blue region), revealing a counter-intuitive benefit of preference opposition.
		\item The decision gap is largest when both members' diversities exceed a threshold and the difference between their aligned preferences is large (bright yellow region). 
	\end{enumerate}
\end{obs}

These observations highlight key dynamics of strategic revelation. 
The first observation shows that when preferences conflict, the members' competitive information sharing leads to a balanced outcome approximating the baseline. 
The second observation captures a critical transition from OPR to BPR-Aligned equilibrium:
In OPR, only Alice reveals information, so the decision gap widens as Bob's diversity increases.
Once Bob's preference loss becomes sufficiently large, he also reveals information, leading to BPR-Aligned equilibrium and a reduced decision gap.
These decision gaps highlight a fundamental limitation of digital representatives in perfectly capturing members' preferences.

\paragraph{Team Preference Loss Comparison}	\label{para: num e comp}
We compare the equilibrium team preference loss $e_T^*$ with the baseline $e_T^B$ in Figure \ref{fig: 2-ne-et-comp}.

\begin{obs}	\label{obs: et comp}
	From Figure \ref{fig: 2-ne-et-comp}, we observe: 
	\begin{enumerate}
		\item The equilibrium team preference loss consistently equals or exceeds the baseline loss, confirming the theoretical optimality of direct preference expression.
		\item With small Alice diversity ($|d_A| = 0.5$), team preference loss follows a non-intuitive pattern: First increasing, then decreasing as Bob's diversity $d_B$ deviates from $0.5$.
		\item With large Alice diversity ($|d_A| = 3$), team preference loss exhibits a complex non-monotonic pattern: Increasing, decreasing, then increasing again as Bob's diversity grows.
	\end{enumerate}
\end{obs}

The first observation confirms Theorem \ref{thm: comp}.
For the second observation, when Alice's diversity is small (blue curve, circle markers), she reveals no information.
In NR equilibrium ($|d_B| < 1$), the team decision equals the GenAI's prior, minimizing the team preference loss $e_T^*$ at $d_B = 0.5$ where they have opposite preferences.
As Bob's diversity increases ($|d_B|>1$), he begins to reveal information, so the team preference loss decreases.

For the third observation, when Alice's diversity is large (orange curve, square markers), she partially reveals information.
In the non-monotonic region ($d_B > 0$), the team preference loss increases initially as Bob's diversity grows ($d_B\in(0, 1.75)$, OPR equilibrium) due to the increasing team decision gap.
Then, the team preference loss decreases once Bob starts revealing information ($d_B \in (1.75, 3)$, BPR-Aligned equilibrium). 
Finally, as Bob's diversity grows further, Alice reduces revelation, increasing the team preference loss ($d_B \in (3,5)$).

\paragraph{Individual Preference Loss Comparison}
Given the complex trends in team preference loss, we examine individual preference loss differences $\Delta e_m \triangleq e_m^* - e_m^B$ in Figure \ref{fig: 2-ne-em-comp}.

\begin{obs}	\label{obs: em comp}
	Surprisingly, the member with a smaller diversity can achieve a lower individual preference loss at equilibrium compared to baseline, despite $e_T^* \ge e_T^B$.
\end{obs}

This counterintuitive result occurs in two scenarios: 
\textit{1)} For Bob with small diversity ($d_B \in (-0.5, 1.75)$): Only Alice reveals information, but her imperfect digital representative inadvertently shifts the team decision closer to Bob's preference.
\textit{2)} For Alice, when her diversity is smaller than Bob's ($d_B \in (4.25, 5)$): In BPR-Aligned equilibrium, Alice strategically reduces her revelation as Bob's diversity increases, obtaining a more satisfying team decision.

\paragraph{Team Total Loss Comparison}	\label{para: num lt comp}
We compare the team total loss between equilibrium $L_T^*$ and baseline $L_T^B$ under different manual costs $C$ in Figure \ref{fig: 2-ne-lt-comp}.

\begin{obs}	\label{obs: lt comp}
	From Figure \ref{fig: 2-ne-lt-comp}, we observe: 
	\begin{enumerate}
		\item Low manual cost ($C = 0.5$): Manual decision-making yields lower team total loss.
		\item High manual cost ($C = 2$): Digital representatives reduce team total loss when preferences align closely with the GenAI's prior.
	\end{enumerate}
\end{obs}

This confirms Theorem~\ref{thm: comp}: Low-cost tasks favor manual decision-making, while high-cost tasks benefit from digital representatives, especially with low preference diversity.

\paragraph{Impact of More Advanced GenAI}	\label{para: num gai}
We investigate the impact of more advanced GenAI (lower trigger price $p$) on team total loss in Figure \ref{fig: 2-ne-beta-comp}.

\begin{obs}	\label{obs: p comp}
	Lower trigger prices enable more diverse members to benefit from digital representatives, reducing team total loss.
\end{obs}

This highlights how advanced GenAI can effectively support diverse preferences in collective decision-making, expanding its practical applications as technology improves.


\section{Conclusion}	\label{sec: conclusion}
This paper investigates the use of GenAI as digital representatives in subjective collective decision tasks.
Our theoretical analysis yields several key insights:
\textit{1)} Conflicting preferences interestingly stimulate greater information revelation at equilibrium; 
\textit{2)} Team decisions made via representatives closely approximate manual baselines when member preferences conflict;
and \textit{3)} While team preference loss never falls below baseline levels, individual members may strategically achieve lower preference losses, revealing complex group dynamics.
These insights provide practical guidance on deploying digital representatives.
Future work should focus on developing mechanisms to mitigate the imperfect preference representation that arises, particularly when preference alignment leads to strategic information withholding.
Examining the model with more comprehensive team settings also deserves further attention.



\begin{ack}
	This work is supported by the National Natural Science Foundation of China (Project 62472367 and Project 62271434), 
	Shenzhen Key Lab of Crowd Intelligence Empowered Low-Carbon Energy Network (No. ZDSYS20220606100601002), 
	Shenzhen Science and Technology Program (Project JCYJ20230807114300001, JCYJ20220818103006012),
	the Shenzhen Stability Science Program 2023, 
	the Shenzhen Institute of Artificial Intelligence and Robotics for Society, 
	and Longgang District Shenzhen's "Ten Action Plan" for Supporting Innovation Projects (No. LGKCSDPT2024002).
\end{ack}



\bibliography{mybibfile}

\end{document}